\documentclass[10pt]{article}
\usepackage{graphicx,color,amsmath}

\newcommand{\figqp}{%
\begin{figure}[htbp]
   \includegraphics[height=1in,clip]{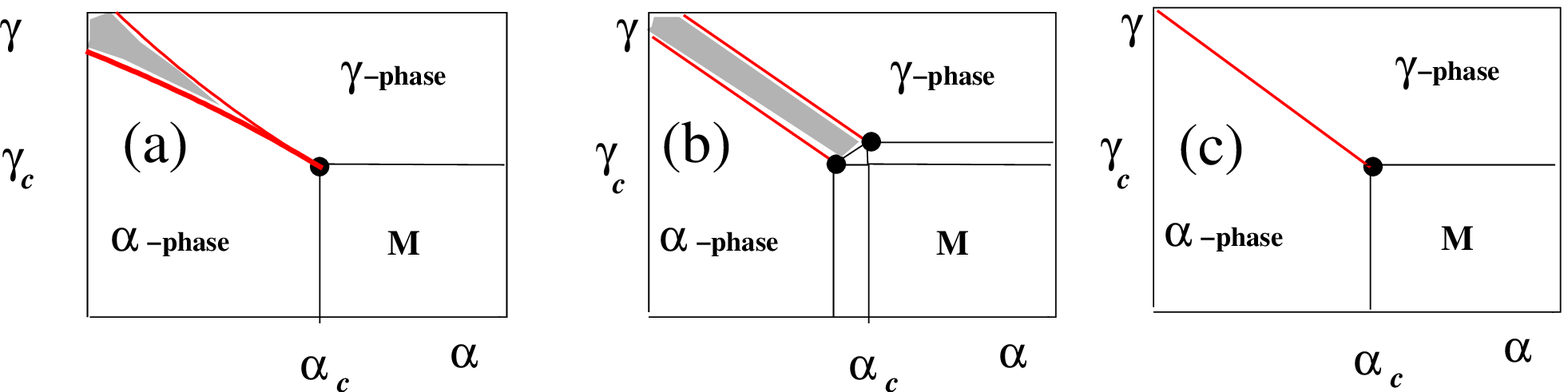}
   \includegraphics[height=1in,clip]{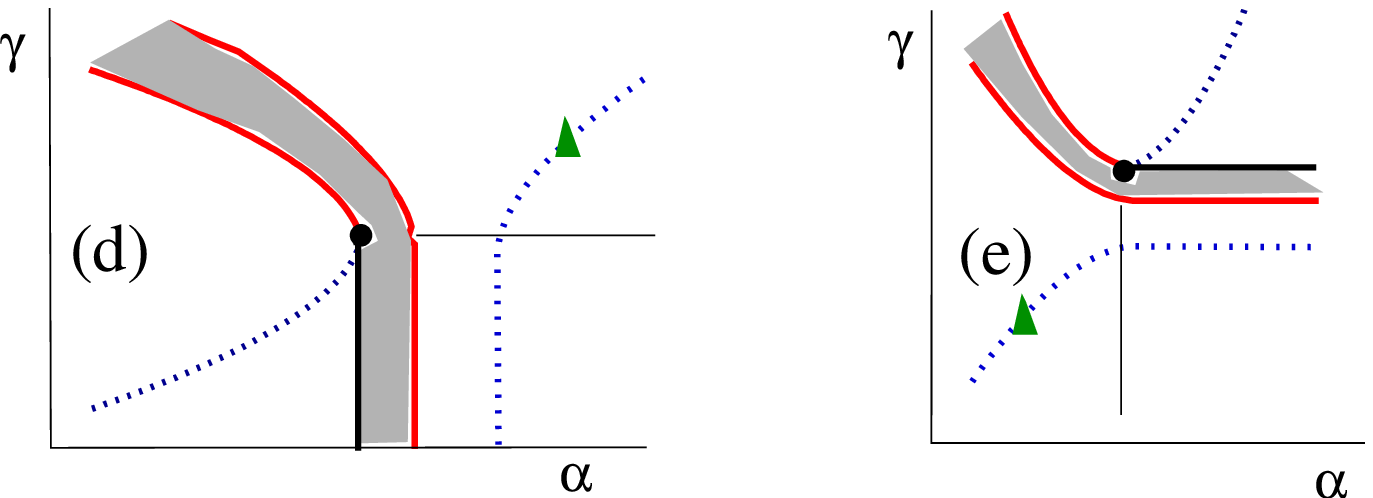}
    \caption{$\alpha-\gamma$ phase diagrams. (a,b,c)  for
      $\rho_L=\rho_m$, (c) for $\rho_L>\rho_M$ and (d) for
      $\rho_L<\rho_M$.  The critical points are represented by filled
      circles.  The shaded region is the shock region. In (a) $q>p$.
      (b) $q=p$ while (c) has no non-conserving part. The MM dual
      lines are shown in (c) and (d) only.  The filled triangles on
      the dual lines represent the point $(\rho_L,\rho_L)$ which is
      always on the dual line.}
\label{fig:qp}
\end{figure}
}

\newcommand{\figtwph}{%
\begin{figure}[htbp]
   \includegraphics[height=1.in,clip]{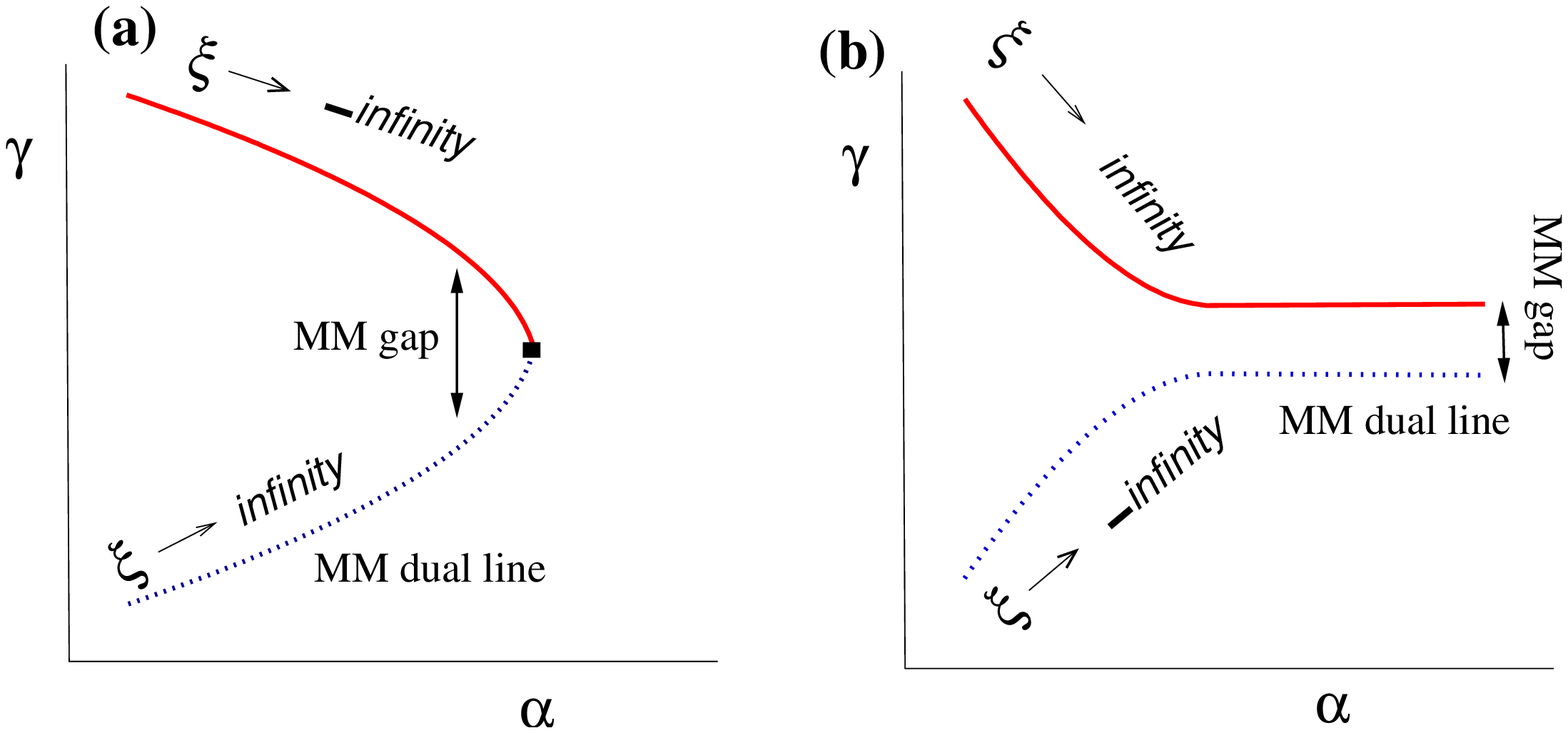}
   \includegraphics[height=1.in,clip]{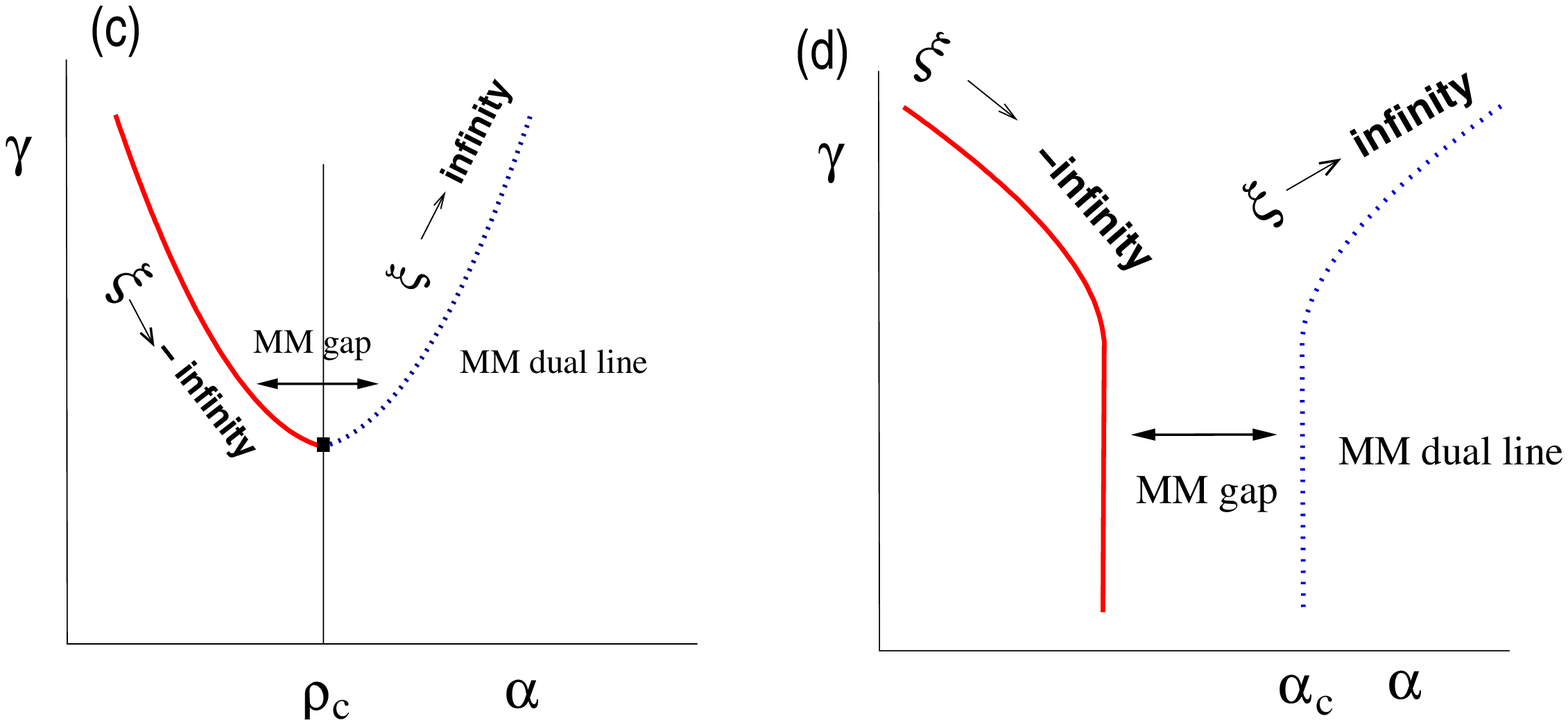}
    \caption{Possible phase boundaries (red thick line) and dual
      lines(blue dotted). (a,b) for the $\alpha$-phase with the shock
      forming at $x=1$ while (c,d) for the $\gamma$-phase with shock
      forming at $x=0$.  In (a) and (c) the intersection of the shock
      phase boundary and the MM dual line produces a critical point
      marked by a filled square. No critical point in (b,d).  The
      $\alpha$,$\gamma$ asymmetry depends on the relative values of
      $\rho_L$ and $\rho_c$.  }
\label{fig:twph}
\end{figure}
}

\newcommand{\tlx}{{\tilde{x}}}
\newcommand{\rht}{\rho_{\rm out}}
\newcommand{\rhn}{\rho_{\rm in}}

\begin{document}

\title{Duality and phase diagram of one dimensional transport}
\author{Somendra M. Bhattacharjee \\ 
Institute of Physics, Bhubaneswar-751005, India\\
Saha Institute of Nuclear Physics, Kolkata - 700064, India}
\maketitle

\begin{abstract}
  The observation of duality by Mukherji and Mishra in one dimensional
  transport problems has been used to develop a general approach to
  classify and characterize the steady state phase diagrams.  The
  phase diagrams are determined by the zeros of a set of
  coarse-grained functions without the need of detailed knowledge of
  microscopic dynamics.  In the process, a new class of nonequilibrium
  multicritical points has been identified.
\end{abstract}

\section{Introduction}
There are many situations that involve transport of particles from one
end to other along a one dimensional track obeying some form of mutual
exclusion\cite{lig}.  Examples are vehicular traffic in a one-lane
road, molecular motors carrying cargo on a track in biological systems
and so on.  Moreover, such simple systems are of importance to develop
an understanding of nonequilibrium steady state phases and phase
transitions.

Many such transport models have now been studied extensively via
analytical methods, meanfield approximations and numerical
simulations, and different types of phases have been
identified\cite{smsmb,smvm,der,gtmb,fst,krug,frey}.  These phases are
represented in phase diagrams in the space of the externally
controllable parameters of the problem, namely, the imposed rates of
injection and withdrawal at the two boundaries.  The fact that the
boundary conditions determine the stable phase diagrams makes these
nonequilibrium problems different from equilibrium ones.

Since phase transitions involve the whole of the system, the generic
behavior, in the large size limit, is expected to be determined by
certain gross overall features.  This is the way equilibrium phase
transitions are analyzed, but, alas, no such general framework is
known for nonequilibrium cases.  Hence the efforts on case by case
studies.  Our aim is to develop a general formulation at least for the
above class of systems.  We show that the generic  features and the
universal properties of the phase diagrams can be obtained from the
structure of a set of $S$-functions and by using a recently
discovered duality\cite{smvm}, without the need of detailed specifics of the
microscopic dynamic rules and interaction.  Still, one microscopic
parameter remains essential for the problem, namely, a small distance
cutoff (e.g.  lattice spacing or some microscopic size etc) which
cannot blindly be set to zero.  The usefulness of the approach is
shown by predicting a new class of nonequilibrium multicritical
points.

\section{Phases and response functions}
As an example\cite{smsmb,smvm,frey}, consider a one dimensional
lattice of $N (\to\infty)$ sites.  Particles are injected at site
$i=1$ at a rate $\alpha$ (i.e probability that a particle is injected
in a short time interval $dt$ is $\alpha dt$) and withdrawn at site
$i=N$ at a rate $1-\gamma$. The particles hop on the lattice as
per preassigned rules, like forbidden multiple occupancy of a site, etc.
In addition, non-conserving processes may allow addition to or
deletion from the track at rates $\omega_a$ and $\omega_d$
respectively, as e.g., exits or feeders in a traffic system for cars
to get out of or into the road, or ``processive''objects in biological
systems falling off the track or getting reattached from the bulk
solution.  The parameters, $K=\omega_a/\omega_d$ and $\Omega=\omega_d
N$ are characteristics of the microscopic dynamics while $\alpha$ and
$\gamma$ are externally imposed.

For a coarse-grained description, the natural variable is the local
density or the space-time dependent average occupation number
$\rho(x,t)$ in continuum ($x\in [0,1]$ by a rescaling of the total
length).  The sensitivity to the boundary concentrations (or rates)
can be measured by the response functions
\begin{equation}
  \label{eq:7a}
  \chi_{\mu}\equiv \frac{\partial M}{\partial \mu}, 
    \ {\rm where}\ \mu=\alpha\  {\rm or}\  \gamma, {\rm and}\  M= \int_0^1\! \!dx\  \rho(x),
\end{equation}
is the steady state spatially averaged density.  Any two points in the
$\alpha-\gamma$ space are said to be in the same phase if they can be
connected by a continuous path along which the density profile or the response
functions change smoothly.  Any point of non-analyticity on a path
defines the location of the phase transition.


The phases observed are, generally, of the following types. (i)
Injection (withdrawal) rate dominated, to be called the $\alpha$-phase
($\gamma$-phase), (ii) a shock phase consisting of piecewise
continuous densities, and (iii) special phases.  In the shock phase,
there is a discontinuity in the density separating an $\alpha$-phase
on one side from a $\gamma$-phase on the other side, while an example
of (iii) is a phase where the current through the system is maximum.
The response functions behave differently in these phases.  In the
$\alpha$-phase, $\chi_{\alpha}\sim O(1)$ but $\chi_{\gamma}=0$, while
in the $\gamma$-phase, it is the other way round.  However, in the
shock phase, both $\chi_{\alpha}$ and $\chi_{\gamma}$ would be nonzero.
For special phases like (iv) in the above list,
$\chi_{\alpha,\gamma}\approx 0$.

\section{Equations for dynamics and steady states: Definitions of $S_i$'s}
In a continuum limit for large $N$ (with lattice spacing $a\ll L=Na$)
the time variation of $\rho(x,t)$ can be written in the form of a
continuity equation, as
\begin{equation}
  \label{eq:8}
  \frac{\partial\rho(x)}{\partial t}+\frac{\partial J_0(x,t)}{\partial
    x} =  S_0(\rho,t),
\end{equation}
where the right hand side is the explicit non-conserving contribution
to the change in density.  The left hand side is in the form of a
continuity equation with $J_0(x,t)$ as the current at the site.  In a
mean field approximation, the current is taken to be an implicit
function of position and time through the density, 
so that  $J_0$ can be split into two parts, 
 \begin{equation}
   \label{eq:9}
  J_0=-\epsilon S_2(\rho) \frac{\partial \rho}{\partial x} + J(\rho(x,t)),
 \end{equation}
with a ``bulk'' contribution $J(\rho)$ determined by the local density
and a term that depends on the derivative of the density (``Fick's law'') 
over the lattice spacing, $\epsilon=a/2N$ being small.  This
$\epsilon$-dependent term is a reminiscent of the interactions in the
neighborhood of a site on the underlying lattice. The form of
$S_2(\rho)$ is determined by the microscopic dynamics, but, for
simplicity, we take $S_2(\rho)=1$ here.  Such a form like Eq. \ref{eq:9}
has recently been
shown by Chakrabarti\cite{chak} to arise naturally in a
renormalization group type approach for transport processes and
failures of fiber bundles.  The fact that there are two fixed points
(viewed as a recursion relation) will be of importance to us also.

In the steady state, the system evolves to a time independent density
profile satisfying
\begin{equation}
  \label{eq:5}
  -\epsilon\frac{d\ }{d x} S_2(\rho) \frac{d \rho}{d x} 
+S_1(\rho)\frac{d \rho}{d x} + S_0(\rho)=0.
\end{equation}
The $S_i$ functions encode the dynamics or specialties of the system.
The density satisfies the boundary conditions $\rho(0)=\alpha$ and
$\rho(1)=\gamma$.
The microscopic rules are taken to be sufficiently smooth to warrant
considerations of smooth and analytic functions only.  These 
restrictions can be relaxed if necessary.

For $\epsilon\to 0$, the ensuing first order equation cannot in
general satisfy the two boundary conditions.  Therefore, the
$\epsilon$ term, eventhough looks innocuous in the bulk limit, is
essential.  It defines a new scale $\tlx= x/\epsilon$ in the problem
and this scale is important for the phase transitions.  E.g., the
discontinuity at a shock will be rounded on a scale of $\tlx$ but
would look sharp on a bigger scale.

\subsection{Nature of $S_0$}
Let us first consider the role and the nature of $S_0(\rho)$.  The
zero of the non-conservation function $S_0(\rho)$ is a special
density.  This is the equilibrium like steady state density
$\rho=\rho_L$, $S_0(\rho_L)=0$, the system would evolve to if all
other dynamics, except this non-conserving one, are switched off.  In
such a situation, the density $\rho_L$ can be obtained from the
extrema of a (free energy like) Lyapunov function, ${\hat{S}}_0(\rho)$
such that $S_0(\rho)=-d{\hat{S}_0}/d\rho$.  For stability of the
state, evaporation is to be preferred in case of excess density (over
$\rho_L$), but adsorption for $\rho<\rho_L$.  This requires
$S_0(\rho)$ to be an odd function or, ${\hat{S}}_0(\rho)$ an even
function of $\rho-\rho_L$.  A simple possible choice is
\begin{equation}
  \label{eq:1}
 {\hat{S}}_0(\rho)=\Omega (\rho-\rho_L)^{2q},\  (q\ge 1),
\end{equation}
with $q=1$ corresponding to a linear form for $S_0$ (the so-called
Langmuir kinetics).  The ``softness''of the state is determined by the
parameter $\Omega$ that controls the width of the well.  Also, the
conserved case is recovered by taking $\Omega\to 0$.  A bistable (or
multi-stable) situation can be obtained for $\Omega <0$ with
additional terms in ${\hat{S}}_0$.

\subsection{Nature of $S_1$}
A zero of $S_1(\rho)$, i.e. $S_1(\rho_m)=0$, is the density at which
the current is an extrema (e.g. a maximum).  From Eq. \eqref{eq:5}, we
see that this is the density where one may afford a non-existence of
the first derivative of the density.  Consequently, a shock, for which
the derivative is not defined (strictly in the $\epsilon\to 0$ limit),
if exists, has to be centered around $\rho=\rho_m$.  If the
dynamics has a particle hole symmetry, then $\rho_m=1/2$.  For
concreteness and simplicity we consider the class of functions
\begin{equation}
  \label{eq:3}
J(\rho)\approx J_{\rm m} - a(\rho-\rho_m)^{2p}, \ (p\ge 1),  
\end{equation}
near the maximum.  There can be cases with more than one zero of
$S_1(\rho)$, which can lead to multiple shocks and more exotic
phenomena. Such cases will be discussed elsewhere.

\subsection{Special cases}
The microscopic dynamic rules determine the values of the two special
densities, $\rho_L, \rho_m$, and the values of $p, q$. However, we do
not require those rules.  We need to distinguish special cases like,
(i) $\rho_L>\rho_m$, (ii) $\rho_L<\rho_m$,and (iii)
$\rho_L=\rho_m$. It is plausible to make a {\it smoothness hypothesis that
the nature of the phase diagram in the $\alpha-\gamma$
(external)-parameter space changes smoothly as the parameters like
$\rho_l, \rho_m$ (determined by the microscopics) are changed unless
there is a special condition}.  Such a condition is $\rho_L=\rho_m$
where the non-conserving processes try to maintain a density at which
the conserved processes can accommodate maximum current.

\section{Boundary layer approach: outer and inner densities}
We adopt the boundary layer approach or the method of matched
asymptotics to handle the two scales in Eq. \eqref{eq:5}. Consider the
case where the bulk density profile $\rho=\rht(x)$, obtained from
Eq. \ref{eq:5} with $\epsilon= 0$, matches the boundary condition
$\rho(x=0)$ = $\alpha$.  But, then, $\rho_{\rm o}$ $\equiv\rht(x=1)$
$\neq \gamma$.  A different density profile $\rhn(\tlx)$ where
$\tlx=(x-x_s)/\epsilon$ $\sim O(1)$, $(x_s=1)$, extrapolates within a
thin ``inner'' region from $\rhn(-\infty)=\rho_{\rm o}$ to
$\rhn(0)=\gamma$. This inner region satisfies $ \left( {\rm see \
Eq. \ref{eq:5}},\ S_1(\rho)={d{\hat{S}}_1(\rho)}/{d\rho}\right)$
\begin{equation}
  \label{eq:2}
-S_2(\rhn) \frac{d \rhn}{d \tlx} +{\hat{S}}_1(\rhn)=0,
\end{equation}
which is equivalent to Eq. \ref{eq:9} with
${\hat{S}}_1(\rho)=J(\rho)-J_0$.  The inner region, to first order in
$\epsilon$, is too thin for the violation of conservation to matter,
so that the current $J_0=J(\rho_{\rm o})$ entering from the bulk
(outer) region remains conserved in the inner layer.  Inference:
{\it The mandatory matching condition requires
${\hat{S}}_1(\rho)$ to have a zero at} $\rho=\rho_{\rm o}$.

\subsection{Zeros as requirements for shocks}
A shock is formed only if the inner solution fails to satisfy the
boundary condition.  This happens if the inner solution saturates at
the other end.  Therefore the {\it minimal requirement for shock
formation is another zero of} ${\hat{S}}_1(\rho)$, so that
\begin{equation}
  \label{eq:6}
  {\hat{S}}_1(\rho)=-(\rho -\rho_{\rm o})(\rho-\rho_{\rm s})\Phi(\rho).
\end{equation}
The first nontrivial case, is therefore a function with
two simple zeros and $\Phi(\rho)=1$.  
The two zeros $\rho_{\rm o}$ and $\rho_s$ correspond to the
two fixed points of Chakrabarti's approach\cite{chak}.
By Rolle's theorem, $\rho_{\rm o} \le \rho_m \le \rho_{\rm s}$.

The inner equation admits two types of solutions, one bounded (B-type)
between $\rho_o$ and $\rho_s$ while the other one shows a divergence
(U-type) with $d\rho/dx \sim -\rho^2$, or more generally, $d\rho/dx
\sim -\rho^{2p}$. 
It is the B-type layers
that matures to a shock but not the U-type.  The inner solution is of
the form ${\cal I}(\tlx/w + \xi)$ with ${\cal I}(z)\to \rho_s$ or
$\rho_{\rm o}$ as $z\to \pm\infty$. Here $w$ is the width of the layer
and $\xi$ gives the location of the center of the layer.  So instead
of the two boundary conditions describing the layer, we may instead
opt for the $w,\xi$ pair.  The center may lie outside the physical
range or may be in an unphysical density range, requiring continuation
of the density and the space beyond the physical range of $[0,1]$.
This continuation helps in getting the general form of the phase
diagram.  The origin is to be called a ``virtual origin'' if it is in
the unphysical region.

\subsection{Shockening transitoin and Mukherji-Mishra dual line}
For a given $\alpha$, as $\gamma$ is changed, two different situations
may arise.  In one, the virtual origin approaches the boundary at
$\tlx=0$ (i.e. $x=1$) and then enters the physical region, eventually
moving to $-\infty$.  In the other situation, the origin remains
virtual and moves to infinity, $\xi\to +\infty$. This is the
Mukherji-Mishra (MM)dual boundary line.

The first case is a thickening of the layer but remaining pinned to
the boundary.  Ultimately as $\xi\to -\infty$, the layer gets released
from the boundary and moves into the bulk.  {\it Or a shock forms}.
So long as the boundary layer stays pinned to the boundary,
$\chi_{\gamma}\sim\epsilon \gamma/{\hat{S}}_1(\gamma) \to 0$ as
$\epsilon\to 0$. In contrast, $\chi_{\alpha}$ is nonzero. The phase,
by definition, is then an $\alpha$-phase. 

The transition of a thin layer to a shock at $\gamma=\rho_s(\rho_{\rm
  o}(\alpha))$ has been called a ``shockening'' transition or a layer
``shockens''.  Beyond this, the layer is separated from the boundary
by a bulk phase (outer solution) of nonzero thickness.  Though the
shockening of the inner layer is a depinning phenomenon at the
boundary, it also signals a bulk phase transition from an
$\alpha$-phase to a shock phase. It is apparent that the shock phase
has both the response functions $\chi_{\alpha,\gamma}\neq 0$.

The symmetry of the two zeros of ${\hat{S}}_1(\rho)$ suggests that
there has to be another line $\gamma=\rho_{\rm o}(\alpha)$ at which
$\xi \to +\infty$.  The boundary region goes from an accumulated to
a depleted region as one crosses this MM line, thereby separating the
shockening (B-type) to nonshockening (U-) type boundary layers.
The MM dual line is purely a boundary transition line, and its existence
is a requirement for shock formation.

\subsection{Two lengthscales}
For $\gamma$ near the two extreme values $\rho_{\rm o}$, $\rho_s$, the
lengthscale $\xi$ can be obtained as $\xi \sim {\cal
  I}^{-1}(\gamma/\rho_X)$ where $X$ stands for ${\rm o}$ or $s$ as
appropriate and $f^{-1}$ is the inverse function of ${\cal I}$
(defined for the inner solution).  Eq.  \eqref{eq:6} suggests, ${\cal
  I}^{-1}$ to be logarithmic implying
$\xi\sim\log\mid\gamma-\rho_X\mid$.  We note here that from the exact
solution of the totally asymmetric exclusion problem (with
conservation), one can associate this dual line ($\alpha=\gamma$) with
$\xi\sim \log\mid \alpha -\gamma\mid$, identical to the result we just
derived.

\figtwph 

The other length scale $w$ can be obtained from various limits of Eq.
\eqref{eq:5}, the lengths differing by constant factors.  From the
asymptotic approach to the limits $\rho_X$, (X=o or s),
$w^{-1}=(\rho_s-\rho_{\rm o})\Phi(\rho_X)$, while for 
$\rho\approx \rho_c$, one gets 
$w^{-1}\sim \mid \rho_{\rm o}-\rho_s\mid/{\hat{S}}_1(\rho_m)$.  
What is important to note is that
for a given $\alpha$, the width is determined by the corresponding
separation of the shockening and the dual line, to be called the
MM-gap. The height $h$ of the shock on the shockening line is also
equal to this MM-gap.

\subsection{Condition for Critical point}
In case the shockening transition line and the dual line intersect,
then the intersection is at $\gamma=\rho_c$ with $w\to\infty$ as
\begin{equation}
  \label{eq:7}
w\sim \mid\rho_{\rm o}-\rho_s\mid^{-(2p-1)} \sim h^{-(2p-1)}.
\end{equation}
Such a divergence is the signature of a critical point.  The bulk
phase transition from the $\alpha$-phase to the shock phase is first
order because at the transition point $h>0$.  On the other hand, the
shock evolves from a zero height at the critical point so that it is a
continuous transition.  In case the two lines do not cross, there will
be no critical point and the lines will span the whole phase diagram,
symmetrically placed around $\gamma=\rho_m$ if $\Phi(\rho)=1$.

\figqp

\subsection{Phase diagrams}
So far we have concentrated on the $\alpha$-phase only.  A similar
analysis can be done for the $\gamma$-phase for which the shock is
formed at $x=0$.  Here again there are two possibilities; the
shockening and the dual lines either intersect at $\alpha=\rho_m$ or
do not intersect but remain on two sides of $\alpha=\rho_m$. All the
four possibilities are shown in Fig. \ref{fig:twph}.  In these
diagrams the lines at $\alpha=\alpha_c$ or $\gamma=\gamma_c$ or both
remain special like the dual lines, representing boundary layer
transitions.

Combining the two, we can now draw the global phase diagram.
Combination of (a) and (c) of Fig. \ref{fig:twph} one gets the type
known for the $\rho_L =\rho_m$ ($K=1$) case of Ref. \cite{smsmb,smvm,frey}.
Similarly, (a) with (d) is known for $\rho_L >\rho_m$, while (b) with
(c) will be the case for $\rho_L <\rho_m$.  For
$\gamma<\gamma_c=\rho_m$, on the $\alpha$-phase side, the U-type
boundary layer renders an effective boundary value $\rho_m$ at $x=1$
and therefore the critical behavior continues for all $\gamma$. The
shock on changing $\alpha$ evolves in height and shifts to $x=0$. On
the $\gamma$-phase side, the response function $\chi_{\gamma}$
undergoes a change on crossing the line $\gamma=\rho_m$, even though
the bulk density distribution changes smoothly.  The line
$\gamma=\rho_m$, like the MM dual lines, indicates a boundary
transition, but in special situations it may develop into a bulk phase
boundary also.  The latter happens as $\rho_L \to \rho_m$.

The shape of the shockening curve near the critical point for $q=1$ is
given by $(\gamma-\gamma_c)\sim \mid\alpha-\alpha_c\mid^{1/2p}$.  For
$\gamma=\gamma_c=\rho_m$, the shock height vanishes on the shock side
as $h\sim \mid\alpha-\alpha_c\mid^{\zeta^{\prime}}$ with
$\zeta^{\prime}=1/(2p+1)$.  Though the $p=1$ case is known in the
literature, we have identified the whole class of multicritical
points.  All the exponents\cite{smvm} associated with the shockening
transition and the critical point can be determined in terms of $q, p$.
These details will be reported elsewhere.

For $\rho_L=\rho_m$, our analysis via the duality yields the nature of
the critical points.  For $q=p$, the phase boundaries are similar to
the $q=p=1$ case.  However for $q>p$, the critical point is at
$\alpha=\rho_m,\gamma=\rho_m$.  We show a new type of phase diagram
for a particular case with $\rho_m=0.5$ in Fig. \ref{fig:qp}.

As one traverses the shock phase from one phase boundary to the other,
the shock position goes from $x=1$ to $x=0$.
Now, if the non-conserving part of the dynamics is removed, the shock
region collapses on to a line as in Fig. \ref{fig:qp}(c).  But the
collapse also means that the shock is uniformly distributed over the
entire length and the density is to be averaged over this distribution
of shocks\cite{pkm}. This yields the linear density profile one knows
from exact solutions\cite{der}. This also shows that the mean field
theory puts a bias towards shock formations, so that a judicious use
is called for in situations where shocks are not expected.

\section{Summary}
In summary, the MM duality theorem can be stated as follows: (a) Every
shockening transition has a dual boundary transition.  (b) If the two
lines (the shockening transition and the dual line) intersect, there
is a critical point.  (c) The nature of the critical point is
determined by the zeros of the $S_i$-functions.  This theorem has been
used to predict the behavior of a class of multicritical points in the
steady state phase diagram of nonequilibrium transport.  Our
conclusion is that all microscopic perturbations need not be relevant
to the phase diagrams.  The microscopic dynamics needs to be analyzed
for the nature of the zeros of the $S_i$ functions and the values of
$q, p$.  Those interactions or rules that change the density
parameters $\rho_{L,m}$ without change in $q, p$, can be grouped into
the same class.  Perturbtions within the class will only make cosmetic
changes in the phase diagram.  We have shown a few examples of
possible multicritical phase diagrams.  New classes can be generated
by including extra features of $S_i$'s.

\end{document}